\documentclass[11pt,twoside]{article}
\usepackage{amssymb}
\usepackage{epsfig}
\usepackage{amsfonts}
\evensidemargin\oddsidemargin
\def \sech{\mathop{\rm sech}\nolimits}
\def \arctanh{\mathop{\rm arctanh}\nolimits}

\begin{document}

\label{xyzt} \leavevmode\vadjust{\vskip -50pt}

{\noindent\sl
Proceedings of the XI Fall Workshop on Geometry and Physics,\\
Oviedo, 2002\\
\rm Publicaciones de la RSME, vol.~xxx,
pp.~\pageref{xxx}--\pageref{endxxx}. } \null\vskip 10mm

\pagestyle{myheadings} \markboth {{\small\sc Jacobi Metric and
Morse Theory of Dynamical Systems}} {{\small\sc A. Alonso
Izquierdo et al.}}

\thispagestyle{empty}

\begin{center}
{\Large\bf Jacobi Metric and Morse Theory of Dynamical

\medskip

Systems}

\vskip 10mm

{\large\bf $^1$A. Alonso Izquierdo, $^1$M.A. Gonz\'alez Le\'on,

$^2$J. Mateos Guilarte and $^2$M. de la Torre Mayado}

\smallskip

{\it $^1$Departamento de Matem\'atica Aplicada, Universidad de
Salamanca }

{\it $^2$Departamento de F\'{\i}sica, Universidad de Salamanca

\smallskip

emails: alonsoiz@usal.es, magleon@usal.es,

guilarte@usal.es, marina@usal.es }
\end{center}

\medskip

\begin{abstract}
\parindent0pt\noindent
The generalization of the Maupertuis principle to second-order
Variational Calculus is performed. The stability of the solutions
of a natural dynamical system is thus analyzed via the extension
of the Theorem of Jacobi. It is shown that the Morse Theory of the
trajectories in the dynamical system is identical to the Morse
Theory of geodesics in the Jacobi metric, even though the
second-variation functionals around the action and the Jacobi
length do not coincide. As a representative example, we apply this
result to the study of the separatrix solutions of the Garnier
System.

\medskip

\it Key words: Maupertuis Principle, Morse Theory.


\end{abstract}

\section{Introduction}

The Jacobi version of the Maupertuis principle establishes that
the dynamics in a natural system can be viewed as geodesic motion
in an associated Riemannian manifold: If $S=\int (T-U) dt$ is the
(natural) action functional of a system defined in a Riemannian
manifold, $(M,g)$, the critical trajectories of $S$ with energy
$E=T+U$ coincide with the extremals (geodesics) of the length
functional $L^J=\int ds$ defined in $(M,h)$, where $h$ is the
Jacobi metric $h=2(E-U) g$.

In this work we analyze whether or not this idea works at the
level of second-order variational calculus. Namely, can the
Hessian operator of a natural dynamical system be viewed as the
geodesic deviation operator of the associated Riemannian manifold?
Early on,it was pointed out that the answer is negative
\cite{Laugwitz} and consequently the stability criterion provided
by the geodesic deviation operator (\lq\lq geometric stability
criterion") does not coincide with the usual criterion based on
analysis of the spectrum of the Hessian operator (\lq\lq dynamical
stability criterion"). We shall show that even in this situation
the Morse theory associated with the dynamical problem is the same
as the Morse theory built on the geodesics of the Jacobi metric.
The crux of the matter is that the Jacobi weak stability criterion
provided by the number of conjugate points crossed by a trajectory
is the same from both points of view.

There are several topics where the geometric criterion has been
used, e.g. study of the chaotic behaviour of Hamiltonian systems
\cite{Casetti} or the non-integrability of dynamical systems
\cite{Kozlov}. The Jacobi-Maupertuis principle has been extended
to the case of Lorentzian manifolds in \cite{Szydlowski2}.

\section{The Maupertuis-Jacobi Principle}

\subsection{Preliminaries and Notation}

Let $M\equiv (M,g)$ be a Riemannian manifold. Geodesics in $M$ are
extremals of the free-action $S_0$ or length functionals $L$ for
any differentiable curve $\gamma:[t_1,t_2]\to M$:
\begin{equation}
S_0[\gamma]=\int_{t_1}^{t_2} \frac{1}{2} \| \dot{\gamma}(t)\|^2 \,
dt;\qquad L[\gamma]=\int_{t_1}^{t_2} \| \dot{\gamma}(t)\| \,
dt\qquad . \label{uno}
\end{equation}
The critical paths are the solutions of the Euler-Lagrange
equations:
\begin{equation}
\delta S_0=0\Rightarrow \nabla_{\dot{\gamma}}\dot{\gamma}=0;\quad
\delta L=0 \Rightarrow
\nabla_{\dot{\gamma}}\dot{\gamma}=\lambda(t) \dot{\gamma}, \quad
\lambda(t)=-\frac{d^2t}{ds^2}\, \left(
\frac{ds}{dt}\right)^2\qquad .\label{dos}
\end{equation}
As a consequence of the invariance of the length functional under
reparametrizations, variational calculus on $L$ leads to equations
where the geodesics are parametrized by an arbitrary parameter $t$
(often called pre-geodesics).

To decide whether a critical path is a local minimum we focus on
the second variation functionals:
\[ \delta^2 S_0=\int_{s_1}^{s_2}
\left\langle \Delta V,V\right\rangle\, ds\, ,\quad \delta^2
L=\int_{s_1}^{s_2} \left\langle \Delta
V^\perp,V^\perp\right\rangle \, ds
\]
\[
\Delta V=-\nabla_{\gamma'}\nabla_{\gamma'} \, V -R(\gamma',V)
\gamma'=-\frac{{\rm D}^2 V}{ds^2}-K_{\gamma'}(V)\qquad .
\]
Here, $V$ denotes a proper variation ($V(s_1)=V(s_2)=0$, see for
instance \cite{Jost} for details) and $V^\perp$ is the component
of $V$ orthogonal to the geodesic. $R$ and $K$ are respectively
the curvature and sectional curvature tensors. Thus, $\Delta$ is
the geodesic deviation operator.

In a natural dynamical system where the configuration space is the
Riemannian manifold $(M,g)$, the dynamics is governed by the
action:
\begin{equation}
S[\gamma]=\int_{t_1}^{t_2} \left( \frac{1}{2} \|
\dot{\gamma}\|^2-U(\gamma(t))\right) \, dt\qquad ,\label{action}
\end{equation}
and the Euler-Lagrange-Newton equations are:
\begin{equation}
\delta S=0\Rightarrow \frac{{\rm D} \dot{\gamma}}{dt}=-{\rm
grad}U\qquad . \label{newton}
\end{equation}
The local stability of a given extremal under proper variations is
determined by the Hessian or second-variation functional:
\begin{equation}
\delta^2 S=\int_{t_1}^{t_2} dt \,  \left( \left\langle \Delta
V,V\right\rangle -H(U) (V,V)\right) =\int_{t_1}^{t_2} dt \,
\left\langle \Delta V-\nabla_V {\rm grad} U,V\right\rangle\qquad .
\label{hessian}
\end{equation}
and we shall denote the differential operator in the quadratic
form (\ref{hessian}) as: $\bar{\Delta}V=\Delta V-\nabla_V{\rm
grad}U$.

\subsection{Geodesics in the Jacobi Metric}

Let $x^i$ be a system of local coordinates in $M$, and let us set
$ds_g^2=\sum_{ij} g_{ij}dx^i \otimes dx^j\equiv \sum_{ij}
g_{ij}dx^i dx^j$. The Jacobi metric $h$ is defined as a conformal
transformation of $g$: $ds_h^2=\sum_{ij} h_{ij}dx^i dx^j$,
$h_{ij}=2(i_1-U)g_{ij}$, where $i_1$ is a constant and the
Riemannian character of $h$ restricts the admissible values for
$i_1$ by means of the inequality $U<i_1$.

Given any two vector fields $X,Y\in \Gamma (TM)$, we shall use the
following notation: $ h(X,Y)=\left\langle X,Y\right\rangle^J$, $
\| X\|^J=\sqrt{\left\langle X,X\right\rangle^J}$, and we shall
write $s$ for $s_h$ for simplicity: $ds\equiv ds_h$.

\medskip

\noindent {\bf Theorem of Jacobi.} {\sl The extremal trajectories
of the variational problem associated with the functional
(\ref{action}) with energy $E=i_1$, are the geodesics of the
manifold $(M,h)$, where $h$ is the Jacobi metric: $h=2(i_1-U) \,
g$.}

The proof of this theorem can be found in several References (see
\cite{Laugwitz} for instance). From an analytic point of view, the
theorem simply establishes that the Newton equations
(\ref{newton}) arising from the action $S$ are tantamount to the
geodesic equations in $(M,h)$:$ \nabla_{\gamma'}^J \gamma'=0$,
$\gamma'=\frac{d\gamma}{ds}$. The equivalence is shown by
performing both a conformal transformation, $h=2(i_1-U)\, g$, and
a reparametrization from the dynamical time $t$ to the arc-length
parameter $s$ in $(M,h)$,
\begin{equation}
\frac{ds}{dt}=2\sqrt{(i_1-U({\gamma}(s)))T}=(i_1-U(\gamma(s)))\qquad
, \label{eqparameter}
\end{equation}
over the solutions  (note that in the domain $U<i_1$ the
reparametrization is well defined).

\section{\lq\lq Geometric Stability\rq\rq    versus \lq\lq
Dynamical Stability\rq\rq}

In the language of Variational Calculus, the Jacobi theorem states
that $\delta S=0\Leftrightarrow \delta S_0^J=0$, or, for geodesics
parametrized with respect to the arc-length, $\delta
S=0\Leftrightarrow \delta L^J=0$. In \cite{articulo} we showed the
following theorems referring to the second-variation functionals:

\noindent {\bf Theorem 1.} {\sl Let $\gamma(t)$ be an extremal of
the $S[\gamma]$ functional, and let $S_0^J[\gamma]$ be the
free-action functional for the Jacobi metric associated with
$S[\gamma]$. The corresponding Hessian functionals satisfy:
\begin{equation}
\delta^2 S_0^J[\gamma]=\delta^2 S[\gamma] +\int_{t_1}^{t_2} dt \,
2 \left\langle \dot{\gamma},\frac{{\rm D}V}{dt}\right\rangle
\left\langle F,V\right\rangle\qquad , \label{theorem1}
\end{equation}
where $F={\rm grad} \ln (2(i_1-U))$.}

\bigskip

\noindent {\bf Theorem 2.} {\sl Let $\gamma$ be an extremal of the
$S[\gamma]$ functional and let $L^J[\gamma]$ be the length
functional of the Jacobi metric associated with $S[\gamma]$. The
corresponding Hessian functionals satisfy:
\begin{equation}
\delta^2L^J[\gamma]=\delta^2 S[\gamma]-\int_{t_1}^{t_2}
\frac{dt}{2(i_1-U)} \left[ \left\langle
\nabla_{\dot{\gamma}}\dot{\gamma},V\right\rangle - \left\langle
\dot{\gamma},\nabla_{\dot{\gamma}}V\right\rangle\right]^2
\label{theorem2}
\end{equation}
}

\noindent Equation (\ref{theorem2}) shows that geodesics of
minimal Jacobi length  $L^J[\gamma]$ are equivalent to stable
solutions of the dynamical system, although the converse is not
necessarily true. Allowing only orthogonal variations,
$V=V^\perp$, (\ref{theorem2}) can be re-written as:
\[
\left. \delta^2S\right|_{V=V^{\bot}}=\delta^2L^J+\int_{s_1}^{s_2}
ds \left( \langle F^J,V^{\bot}\rangle^J\right)^2
\]

The proofs of these two theorems are based on the behaviour of
the covariant derivatives and the curvature tensor under
reparametrizations and conformal transformations of the metric
tensor.

\section{Jacobi Fields and Morse Theory}

A basic tool for the analysis of the stability properties of a
given extremal of a functional is the study of the Jacobi fields:
the elements of the Kernel of the second-variation or Hessian
operator. Direct computation of the Jacobi fields in a given
problem is in general a very complicated task: one needs to solve
the Euler-Lagrange equations, calculate the second-variation
operator and, finally, determine the Kernel. Fortunately, there
exists a short-cut if one knows a parametric family of extremals
(see, for instance, \cite{Giaquinta} for details):

\medskip

\noindent {\bf Proposition.} {\sl If $\gamma=\gamma(t,a)$ is a
family of extremals of $S[\gamma]$ characterized by the value of
the real parameter $a$, then the vector field $\frac{\partial
\gamma}{\partial a}$ is a Jacobi field.}

Knowledge of the Jacobi fields allows us to calculate the
conjugate points along an extremal $\gamma$, and it is thus
possible to apply the Jacobi criterion for weak minimizers of a
functional which counts the number of conjugate points crossed by
the extremal as a measure of the instability.

\medskip

Morse Theory establishes a link between the topology of a manifold
and the critical-point structure of the functions defined in the
manifold \cite{Morse}. We shall deal with Morse Theory in spaces
of closed paths with a fixed point in Riemannian manifolds. The
aim is the application to the separatrix trajectories in the
Garnier System. A formulation of this theory \`a la Bott
\cite{Bott} is as follows:

Let $ \Omega M=\{ \gamma:S^1\to M\, /\, \gamma(0)=\gamma(1)=m_
0\}$ be the loop space in $M$ with a fixed base point. Then, some
of the topological properties of $\Omega M$ are codified in the
Poincar\'e series $P_t\left( \Omega M^n
\right)=\sum_{k=0}^{\infty} b_k\, t^k$, where $b_k={\rm dim} \,
H_k\left( \Omega M, {\Bbb R}\right)$ are the corresponding Betti
numbers.

For any functional $S$ defined on $\Omega M$, we define the Morse
series as ${\cal M}_t\left( S\right) =\sum_{N_c}\, P_t\left(
N_c\right)\, t^{\mu(N_c)}$, where the sum is over the critical
manifolds $N_c$, formed either by a single isolated critical path
or by a continuous set of critical paths. $\mu(\gamma_c)$, the
Morse index, is the dimension of the subspace where the Hessian
operator of the functional $S$, $\Delta$, along $\gamma_c$ is
negative definite.

The Morse inequalities, ${\cal M}_t(S)\, \geq\, P_t\left( \Omega
M\right)$, tell us that the topology of $\Omega M$ forces the
existence of most of the extremals of the functional $S$.

The Morse index can be computed using the {\bf Morse index
Theorem}: {\sl The Morse index of a critical path $\gamma_c$ is
equal to the number of conjugate points to the base point crossed
by $\gamma_c$ counted with multiplicity.} Note that this theorem
is equivalent to the Jacobi criterion for weak minimizers.

From this point of view, knowledge of a one-parametric family of
solutions $\gamma(t; a)$ of the dynamical problem (\ref{action})
informs us about the geodesics $\gamma(s; a)$ in $(M,h)$. There is
a one-to-one correspondence between the trajectories of the
dynamical system and the geodesics of the Jacobi metric, which
only differ by proper reparametrizations. Therefore, the two kinds
of extremals share all  the conjugate points, and the structure of
the critical points of $S$ and $L^J$ is the same. The Morse series
coincide, ${\cal M}(S)={\cal M}(L^J)$, and the Morse theories
built from the trajectories in $(M,g)$ and the geodesics in
$(M,h)$ are identical.

\section{Separatrix trajectories in the Garnier System}

As an application, we shall analyze the separatrix trajectories in
the Garnier System \cite{Garnier}. The configuration space is
$({\Bbb R}^N,\delta_{ij})$ and the action functional reads:
\begin{equation}
S=\int dt  \left\{ \frac{1}{2} \| \dot{\vec{q}} \|^2+\frac{1}{2}
\left( \vec{q}\cdot \vec{q}-1\right)^2 +\sum_{i=1}^N\frac{1}{2}
\sigma_i^2q_i^2\right\}\qquad .\label{1}
\end{equation}
Here $\sigma_i$ are constants, such that $\sigma_i\neq \sigma_j$,
$\forall i\neq j$, and the Euler-Lagrange equations are:
\begin{equation}
\frac{d^2q_i}{dt^2}=2 q_i \left( \vec{q}\cdot \vec{q}-1\right)
+\sigma_i^2q_i,\qquad i=1,2,\dots,N\qquad .\label{ecuacionestau}
\end{equation}

The system is completely integrable and, in fact, Garnier found
all the periodic solutions in terms of hyperelliptic functions. In
\cite{nosotrosNonli} we computed the separatrix solutions lying on
the boundary between bounded and unbounded motion . We shall focus
in the $N=2$ case for simplicity (in \cite{articulo} the very rich
$N=3$ case is analyzed thoroughly).

For any path $\vec{q}(t)=(q_1(t),q_2(t))$ in $({\Bbb
R}^2,\delta_{ij})$, the Lagrangian ${\cal L}={\cal L}(q_1,q_2,$ $
\dot{q}_1,\dot{q_2})$ is:
\begin{equation}{\cal
L}=\frac{1}{2}\left( \dot{q}_1^2+\dot{q}_2^2 \right) +\frac{1}{2}
\left( q_1^2+q_2^2-1\right)^2+\frac{\sigma^2}{2} q_2^2\qquad .
\label{mstb7}
\end{equation}
With no loss of generality, we have set $\sigma_1=0$,
$\sigma_2=\sigma$. We shall only deal with the case $\sigma<1$,
because the separatrix solutions in this regime have a richer
structure. In fact $\sigma>1$ generates a situation where the
separatrices are reduced to a single singular solution. The system
is not only completely integrable, but Hamilton-Jacobi separable
in Jacobi elliptical coordinates: $q_1^2=\sigma^{-2} ( 1-\mu_1) (
1-\mu_2)$, $q_2^2=\sigma^{-2}(\bar{\sigma}^2-\mu_1) (
\bar{\sigma}^2-\mu_2)$, where $\bar{\sigma}^2=1-\sigma^2$. With
this change, ${\Bbb R}^2$ is mapped into the infinite
parallelogram determined by the inequalities: $
-\infty<\mu_1<\bar{\sigma}^2<\mu_2<1$.

The choice of the separatrix trajectories is achieved by setting
the values of the two invariants in involution to zero:
$i_1=i_2=0$. $i_1$ denotes the energy first-integral and $i_2$ is
the second first-integral of the system, functionally independent
of $i_1$ (see for instance \cite{Guilarte} for details). This
choice confines the dynamics to the finite parallelogram
${\bar{\bf P}}(0)\equiv 0\leq \mu_1<\bar{\sigma}^2<\mu_2<1$ (in
Cartesian coordinates it is the area bounded by the ellipse
$q_1^2+\frac{q_2^2}{\bar{\sigma}^2}=1$).

In elliptical coordinates the Euclidean metric $\delta_{ij}$
becomes : $g_{12}=0=g_{21}$,
$g_{11}=\frac{-(\mu_1-\mu_2)}{4(\mu_1-1)(\mu_1-\bar{\sigma}^2)}$,
$g_{22}=\frac{-(\mu_2-\mu_1)}{4(\mu_2-1)(\mu_2-\bar{\sigma}^2)}$.
The Christoffel symbols are obtained from
\begin{eqnarray*}
&\Gamma_{11}^1=
\frac{1}{2(\mu_1-\mu_2)}-\frac{2\mu_1-(1+\bar{\sigma}^2)}{2
(\mu_1-1)(\mu_1-\bar{\sigma}^2)}; \ \Gamma_{12}^1=
\frac{-1}{2(\mu_1-\mu_2)};\  \Gamma_{22}^1=
\frac{(\mu_1-1)(\mu_1-\bar{\sigma}^2)}{2(\mu_2-1)(\mu_2-\bar{\sigma}^2)
(\mu_1-\mu_2)}
\end{eqnarray*}
Replacing 1 by 2, $\Gamma_{22}^{2}$ is equivalent to
$\Gamma_{11}^{1}$, and the same happens between $\Gamma_{12}^2$,
$\Gamma_{11}^2$, and $\Gamma_{12}^1$, $\Gamma_{22}^1$. The rest of
the symbols are trivially related to these by symmetries or are
null.

The Jacobi metric associated with the system, with $i_1=0$, is
$h=\frac{1}{\mu_2-\mu_1} ( - (\mu_1^3-\bar{\sigma}^2\mu_1^2)+
(\mu_2^3-\bar{\sigma}^2\mu_2^2)) g$, and the Christoffel symbols
can be obtained by direct calculation (\cite{articulo}).

\subsection{Singular Solutions}

There are two singular solutions to the system, and these live on
the border of the parallelogram $\bar{\bf P}(0)$.

Taking $q_2=0$, the equations of the motion can be easily
integrated, even in Cartesian coordinates, to find the solutions:
$\vec{q}(t)=(\bar{q}_1(t),0)=(\pm \tanh (t-t_0),0)$. In elliptical
coordinates, these solutions lie on two edges of the border of
$\bar{\bf P}(0)$: I, $\vec{\mu}_{\rm I}(t)= (1- \tanh^2 (\pm
t),\bar{\sigma}^2)$, for $t\in (-\infty ,\arctanh (\mp \sigma)]
\sqcup [\arctanh (\pm \sigma),\infty )$, and II, $\vec{\mu}_{\rm
II}(t)= (\bar{\sigma}^2,1- \tanh^2 (\mp t))$ for $t\in [\arctanh
(\mp \sigma),$ $\arctanh (\pm \sigma)]$ (taking $t_0=0$).

 The geodesic equations of the Jacobi metric reduced to the
$q_2=0$ orbit become a single non-trivial differential equation,
which is solved by a cubic algebraic equation :
\begin{equation}
q_1''+\frac{2q_1}{q_1^2-1} q_1'^2=0\Rightarrow
\bar{q}_1-\frac{\bar{q}_1^3}{3}=\pm s\qquad .\label{jacobitk1}
\end{equation}
$(t_1,t_2)=(-\infty,\infty)$ has been reparametrized to
$[s_1,s_2]=\left[-\frac{2}{3},\frac{2}{3}\right]$. The explicit
solution of the cubic (\ref{jacobitk1}) is
\begin{equation}
q_1(s)=\bar{q}_1(s)= -\cos \frac{\theta}{3}+\sqrt{3} \sin
\frac{\theta}{3}\qquad ,\label{q1tk1jac}
\end{equation}
where $\theta=\arctan \frac{\sqrt{4-9s^2}}{-3s}$ .

\begin{figure}[htbp]
\begin{center}
\epsfig{file=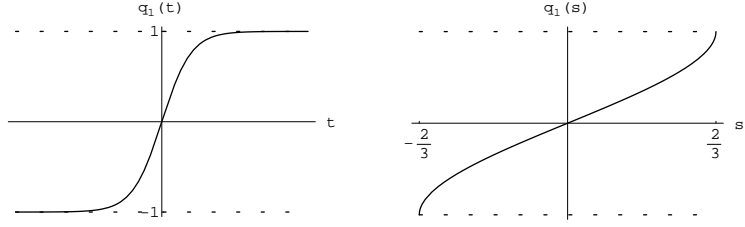,height=3cm}
\end{center}
\caption[Figura]{\small $q_2=0$ singular solution as a function of
the dynamical time $t$ and as a geodesic parametrized by the
arc-length parameter of the Jacobi metric $s$.}
\end{figure}

There exists a second type of singular solutions living on the
edge $\mu_1=0$, which also belongs to the border of $\bar{\bf
P}(0)$. This edge becomes the ellipse
$q_1^2+\frac{q_2^2}{\bar{\sigma}^2}=1$ in Cartesian coordinates.
The solutions are either $ \vec{q}(t)=(\tanh (\pm \sigma
t),\bar{\sigma}\sech (\pm \sigma t))$ in ${\Bbb R}^2$, or
$\vec{\mu}(t)=(0 ,1-\sigma^2 \tanh^2 \sigma t)$ in the elliptic
plane.

The geodesic equations of the Jacobi metric reduced to the
$q_1^2+\frac{q_2^2}{\bar{\sigma}^2}=1$ orbit also become a single
non-trivial differential equation, which is solved by a cubic
algebraic equation :
\begin{equation}
q_1''-\frac{2\sigma^2q_1}{1-\sigma^2q_1^2}q_1'^2=0\, \Rightarrow
\sigma \left( \bar{q}_1-\frac{\sigma^2 \bar{q}_1^3}{3}\right) =
s\qquad . \label{jacobitk2}
\end{equation}
The explicit solution of the cubic together with the elliptic
orbit afford the geodesic:
\begin{eqnarray*}
&\bar{q}_1(s)= -\frac{1}{\sigma}\cos
\frac{\theta}{3}+\frac{\sqrt{3}}{\sigma} \sin \frac{\theta}{3},\
\bar{q}_2(s)= \frac{\bar{\sigma}}{\sigma} \sqrt{ - 2+ \sigma^2 +
\cos\frac{2\theta}{3}+\sqrt{3}\sin\frac{2\theta}{3}}
\end{eqnarray*}
from $s_1=-\sigma (1-\frac{\sigma^2}{3})$ to $s_2=\sigma
(1-\frac{\sigma^2}{3})$.

\begin{figure}[htbp]
\begin{center}
\epsfig{file=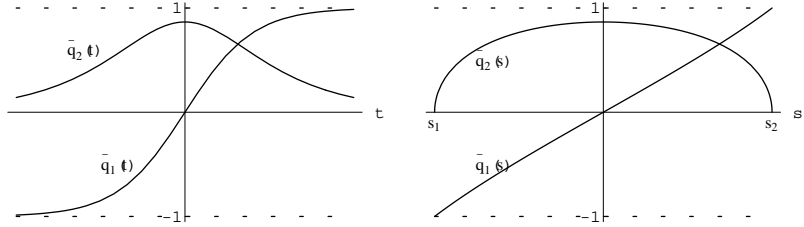,height=3cm}
\end{center}
\caption[Figura]{\small Singular solution over the ellipse as a
dynamical solution and as a geodesic.}
\end{figure}

\subsection{General Solution}
The Hamiltonian associated with the Lagrangian (\ref{mstb7}), in
elliptical coordinates is of the St\"ackel type:
\begin{eqnarray}
H=&&\frac{1}{2(\mu_1-\mu_2)}
\left(-4(\mu_1-1)(\mu_1-\bar{\sigma}^2)\pi_1^2
-4(1-\mu_2)(\mu_2-\bar{\sigma}^2)\pi_2^2-\right. \nonumber \\
&&\left. -(\mu_1^3-\bar{\sigma}^2\mu_1^2)+
(\mu_2^3-\bar{\sigma}^2\mu_2^2)\right) \label{mstbham}
\end{eqnarray}
and the integration of the Hamilton-Jacobi equation, for
separation constants equal to zero, \cite{Guilarte}, leads to the
equation of the orbits,
\begin{eqnarray}
&&\left( \left|
\frac{\sqrt{1-\mu_1}-\sigma}{\sqrt{1-\mu_1}+\sigma} \right| \cdot
\left| \frac{\sqrt{1-\mu_1}+1}{\sqrt{1-\mu_1}-1} \right|^\sigma
\right)^{{\rm sign}(\pi_1)}\cdot\nonumber\\ && \left( \left|
\frac{\sqrt{1-\mu_2}- \sigma}{\sqrt{1-\mu_2}+\sigma} \right| \cdot
\left| \frac{\sqrt{1-\mu_2}+1}{\sqrt{1-\mu_2}-1} \right|^\sigma
\right)^{{\rm sign}(\pi_2)} = e^{2\sigma\bar{\sigma}^2 a} \qquad ,
\label{mstbtray}
\end{eqnarray}
and the equation of the temporal dependence,
\begin{equation}
\left| \frac{\sqrt{1-\mu_1}-\sigma}{\sqrt{1-\mu_1}+\sigma}
\right|^{\bar{\sigma}^2 {\rm sign}(\pi_1)}\cdot \left|
\frac{\sqrt{1-\mu_2}-\sigma}{\sqrt{1-\mu_2}+\sigma}
\right|^{\bar{\sigma}^2 {\rm sign}(\pi_2)}\, =\, e^{2(t+t_0)
\sigma}\qquad .\label{mstbtemp}
\end{equation}
Here $a$ is a constant that parametrizes the different orbits and
$t_0$ is another constant coming from the time-translation
invariance.

At the limit $a\to \pm \infty$, equation (\ref{mstbtray}) reduces
to the singular solution $q_2=0$ together with the singular
solution living on the``ellipse''. Thus, identifying $a=-\infty$
with $a=\infty$, we see that all the orbits are parametrized by a
periodic parameter $a$ : the space of all orbits is the $S^1$
circle.

\begin{figure}[htbp]
\begin{center}
\epsfig{file=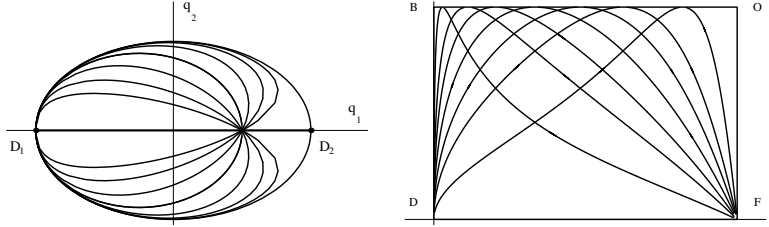,height=3cm}
\end{center}
\caption[Figura]{\small Graphics of the orbits (\ref{mstbtray}) in
the Cartesian plane and in the elliptical one. The graphics were
obtained by numerical calculations.}
\end{figure}

A straightforward application of the Hamilton-Jacobi procedure to
the free-Hamiltonian, $H^J=\frac{1}{2} \left( h^{11}
\tilde{\pi}_1^2+h^{22} \tilde{\pi}_2^2\right)$ ,
$\tilde{\pi}_j=\frac{d\mu_j}{ds}$, provides all the geodesics of
the Jacobi metric with $i_1=0$.

The equation giving the orbits is exactly (\ref{mstbtray}), and
the arc-length dependence is determined by the expression:
\begin{equation}
s+s_0=-(-1)^{{\rm sign}(\pi_1)}\frac{\mu_1+2}{3}
\sqrt{1-\mu_1}-(-1)^{{\rm sign}(\pi_2)}\frac{\mu_2+2}{3}
\sqrt{1-\mu_2}\label{sdependence}
\end{equation}

\subsection{Jacobi fields}

In \cite{nosotrosNonli} the Jacobi fields are calculated on the
trajectories. Here, we closely follow the same idea to compute the
Jacobi fields on the geodesics, which are the solutions of the
equations (\ref{mstbtray})-(\ref{sdependence}). Thus, we have a
two-parametric family $\vec{\mu}(s;a,s_0)$ of geodesics, and,
accordingly, two Jacobi fields: $\frac{\partial
\vec{\mu}}{\partial s_0}$ and $\frac{\partial \vec{\mu}}{\partial
a}$; see Section \S 2.2 . The first Jacobi field is tangent to the
geodesic because it obeys the invariance under $s$ translations.
We focus on the computation of the orthogonal Jacobi field:
$J=\frac{\partial \vec{\mu}}{\partial a}$. An explicit formula for
the geodesic $\vec{\mu}(s;a,s_0)=(\bar{\mu}_1,\bar{\mu}_2)$ in
terms of known analytical functions is not available, but we
implicitly derive the system (\ref{mstbtray})-(\ref{sdependence})
with respect to $a$ to obtain a linear system in the components of
$J$:
\begin{eqnarray}
&&\frac{-(-1)^{\alpha_1}}{\bar{\mu}_1
\sqrt{1-\bar{\mu}_1}(\bar{\sigma}^2-\bar{\mu}_1)} \,
\frac{\partial \bar{\mu}_1}{\partial a}-
\frac{(-1)^{\alpha_2}}{\bar{\mu}_2
\sqrt{1-\bar{\mu}_2}(\bar{\sigma}^2-\bar{\mu}_2)} \,
\frac{\partial \bar{\mu}_2}{\partial a}=2\label{sistema1}\\
&&\frac{(-1)^{\alpha_1}\bar{\mu_1}}{\sqrt{1-\bar{\mu}_1}} \,
\frac{\partial \bar{\mu}_1}{\partial a}
+\frac{(-1)^{\alpha_2}\bar{\mu}_2}{ \sqrt{1-\bar{\mu}_2}} \,
\frac{\partial \bar{\mu}_2}{\partial a}=0\label{sistema2}
\end{eqnarray}

The Jacobi field that solves (\ref{sistema1})-(\ref{sistema2}) is:
\begin{equation}
J=j(\bar{\mu}_1,\bar{\mu}_2)\, \left( (-1)^{\alpha_1} \bar{\mu}_2
\, \sqrt{1-\bar{\mu}_1}  \, \frac{\partial}{\partial \mu_1}+
(-1)^{\alpha_2}\bar{\mu}_1\,  \sqrt{1-\bar{\mu}_2}  \,
\frac{\partial}{\partial \mu_2}\right)\qquad , \label{Jacobifield}
\end{equation}
where
\[
j(\bar{\mu}_1,\bar{\mu}_2)=\frac{2\bar{\mu}_1\bar{\mu}_2
(\bar{\mu}_1-\bar{\sigma}^2)(\bar{\mu}_2-\bar{\sigma}^2)}{(\bar{\mu}_1
-\bar{\mu}_2) (\bar{\mu}_1^2+\bar{\mu}_2^2+\bar{\mu}_1
\bar{\mu}_2-(\bar{\mu}_1+\bar{\mu}_2) \bar{\sigma}^2)},\quad
\alpha_1,\alpha_2=0,1
\]

$J$ is zero both at the starting point D and at the focus F of the
ellipse, and, henceforth, the focus is a conjugate point of D of
multiplicity 1.

\subsection{The Morse series}

Taking $M=\bar{\bf P}_2(0)$; $\Omega M={\cal C}$, we have the
following critical point structure for the geodesics that start
and end at the point D. Firstly, the point D itself is a possible
geodesic with a Morse index 0. Secondly, all members of the family
determined by the general solution, $\vec{\mu}(s;a,s_0)$ cross the
focus. Bearing in mind that $P_t(S^1)=(1+t)$, the family
contributes to the Morse series with $(1+t) t$. Iterating two of
the solutions, $\vec{\mu} \sharp \vec{\mu}$, D is a conjugate
point to itself and the contribution to the Morse series is
$(1+t)t^3$. More iterations tell us that ${\cal M}_t( L^J|_{\cal
C}) = 1+(1+t) t+(1+t) t^3+\dots=\frac{1}{1-t}$. The Morse series
is equal to the Poincar\'e series of the space of closed geodesics
in $S^2$: ${\cal M}_t\left(\left. S\right|_{\cal C}\right)
=P_t\left( \Omega S^2\right)$.

In \cite{nosotrosNonli} we calculated the Morse series for the
$N=3$ case. The result is: ${\cal M}_t\left(\left.
L^J\right|_{\cal C}\right)=P_t\left( {\cal C}\right)
=\frac{1}{1-t^2}= P_t\left( \Omega S^3\right)$.

\end{document}